\documentclass[12pt]{article}
\usepackage[dvips]{color}
\usepackage{mathtext}
\usepackage[koi8-r]{inputenc}
\usepackage{graphicx}
\usepackage[figuresright]{rotating}
\usepackage{amssymb,eucal}
\usepackage{cite}

\newcommand{\beq}{\begin{equation}}
\newcommand{\eeq}{\end{equation}}
\newcommand{\beqn}{\begin{eqnarray}}
\newcommand{\eeqn}{\end{eqnarray}}

\date{}
\begin{document}
\title{The near neutrino detector for the  T2K experiment}

\author{Yury Kudenko\thanks{{\it Email address:} kudenko@inr.ru} \\
~\\
Institute for Nuclear Research RAS \\
60 October Revolution Pr. 7A, 117312 Moscow, Russia \\
~\\
{\it Representing the T2K Collaboration}}
\maketitle
\begin{abstract}
The T2K experiment is a second generation 
long baseline neutrino oscillation experiment designed as a sensitive 
search for $\nu_e$ appearance. 
The  T2K near neutrino detector complex is located 280 meters from the pion production target 
and will measure both neutrino beam properties close to the production point and 
interaction cross sections.   The  main design
features, test results and  status of these detectors    are 
presented. 
\end{abstract}


\section{Introduction}
\label{sec:intro}
In recent years, the atmospheric~\cite{sk}, solar~\cite{solar}, 
reactor~\cite{kamland}, and accelerator~\cite{k2k, minos_first} experiments 
have provided convincing evidence of neutrino oscillations and 
therefore have demonstrated that neutrinos have non--zero masses.  This phenomenon 
is the first clear example of new physics beyond the Standard Model which 
assumes neutrinos are massless particles.  Three 
generation neutrino oscillations are described by six independent parameters: 
three mixing angles ${\rm sin^2\theta}_{12}, 
{\rm sin^2\theta}_{23},  {\rm sin^2\theta}_{13}$,  two mass-squared 
differences $\Delta m^2_{21} = m^2_2 - m^2_1$ and 
$\Delta m^2_{23} = m^2_3 - m^2_2$, and one complex phase $\delta$. 
Both mass differences and two mixing 
angles ($\theta_{12}$ and $\theta_{23}$) are measured. The mixing angle 
$\theta_{13}$ was found to be small and only an upper limit was 
obtained~\cite{chooz,k2k_theta_13}. Presently nothing is known 
about the CP violating Dirac phase $\delta$. The near--future neutrino
oscillation experiments will be focused on the measurements of the unknown 
neutrino  parameters: $\theta_{13}$, mass hierarchy, and  
$\delta$. Another important goal of these experiments is  to 
measure the known mixing parameters more precisely.

\section{Principles of T2K}
\label{sec:t2k}
\begin{figure}[h]
\centering\includegraphics[width=14cm,angle=0]{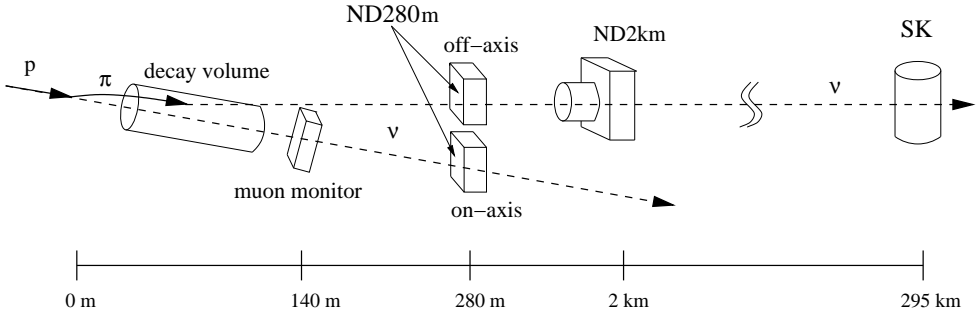}
 \caption{General layout  of the T2K experiment. The basic elements are 
 the neutrino beam line, muon monitor, near neutrino detector at 280 meters  
 from the pion production target, and the far neutrino detector 
 SuperKamiokande. Possible future 2km near detector  is also shown.}
\label{fig:t2k_setup}
\end{figure}

The T2K (Tokai--to--Kamioka) experiment~\cite{t2k}   will use a 
high intensity  off--axis neutrino 
beam generated by a 50 GeV  (initially 30 GeV) proton beam at  JPARC 
(Japan Proton Accelerator Research Complex), SuperKamiokande  as a far 
neutrino 
detector, and  a set of dedicated neutrino detectors    located 
at a distance of 280 m from the pion production target  to
measure the properties of the unoscillated neutrino beam.  The schematic view 
of the T2K setup is shown in Fig.~\ref{fig:t2k_setup}.
 The first phase of 
 T2K   has two main goals: a sensitive 
measurement of ${\rm\theta}_{13}$  and  a more accurate determination of the 
parameters ${\rm sin^22\theta}_{23}$ and $\Delta m^2_{23}$ than any previous
experiment. 

The probability of $\nu_{\mu}$ transition to  $\nu_e$  can be approximately 
given by
\begin{equation}
P(\nu_{\mu}\to \nu_e) \approx 4{\rm cos}^2\theta_{13}{\rm sin}^2\theta_{13}
{\rm sin}^2\theta_{23}{\rm sin}^2\Bigl
(\frac{1.27\Delta m^2_{13}({\rm eV}^2)L({\rm km})}{E_{\nu}({\rm GeV})}\Bigr),
\label{eq:p(numu-nue)}
\end{equation}
where $L$ is the $\nu$ flight distance, and $E_{\nu}$ is the neutrino energy.
It follows from this expression, the maximum sensitivity to the
$\nu_{\mu}\to \nu_e$ transition is expected around the oscillation maximum 
for $\Delta m_{13} \simeq \Delta m_{23} = \Delta m_{atm} \simeq 2.5\times
10^{-3}$. Based on this value, the neutrino peak energy in T2K should be  tuned 
to  $\leq 1$ GeV to
maximize the sensitivity for muon neutrino 
oscillations  for a baseline of 295 km. 

\begin{figure}[h]
\centering\includegraphics[width=10cm,angle=0]{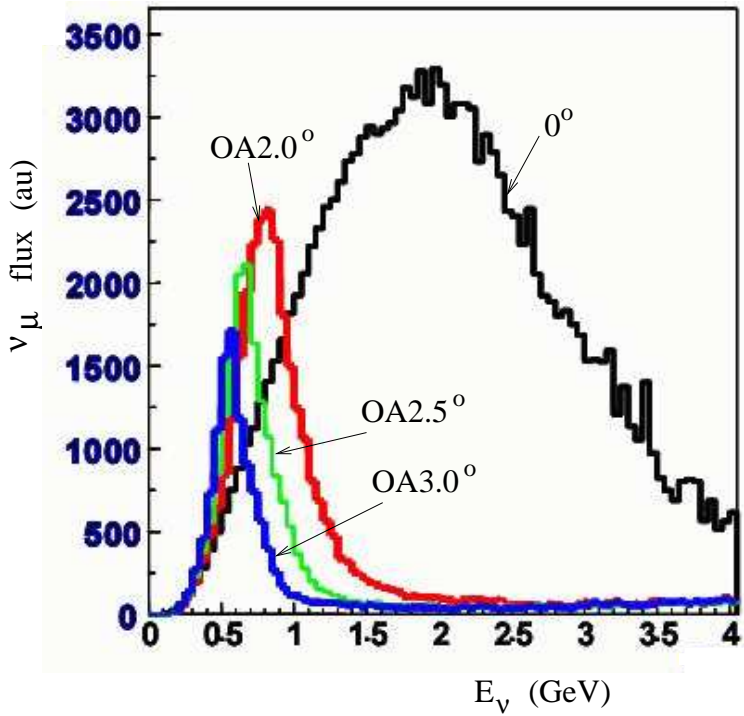}
\caption{Neutrino energy spectra at $0^{\circ}$ and  different 
off-axis angles.}
\label{fig:t2k_nu_beam}
\end{figure}

T2K will adopt an 
off--axis beam configuration in which neutrino energy is almost independent of 
pion energy and quasi-monochromatic neutrino spectrum can be achieved. 
The neutrino beam  is produced 
from pion decays in a 94 m  decay tunnel filled with 1 atm He gas at an
angle of 
$2.5^{\circ}$  with respect to the proton beam axis, providing a  narrow neutrino 
spectrum  with mean neutrino energies from 0.7 to  0.9 GeV, as shown in 
Fig.~\ref{fig:t2k_nu_beam}. 
The high energy tail is considerably reduced at $2.5^{\circ}$ in 
comparison with
the standard on-axis wide-band beam. This minimizes the neutral 
current $\pi^0$ background  in the $\nu_e$ appearance search.
Moreover, the intrinsic contamination of $\nu_e$'s from muon and
kaon decays  is expected to be about 0.4\% around the peak energy.

To achieve T2K  goals,  precise measurements 
of the   neutrino flux, spectrum and 
interaction cross sections are needed. For these purposes, the near detector 
complex (ND280)~\cite{nd280}  will 
be built at a distance of 280 m from the target along the line defined by the 
average pion decay point and  SK (see Fig.~\ref{fig:t2k_setup}). 
This complex has two detectors: an on-axis detector (neutrino beam
monitor)    and  an off--axis detector. Physics 
requirements for ND280  
can be briefly summarized as 
follows: the absolute energy scale of the neutrino spectrum  must
be calibrated with 2\% precision, and the neutrino flux monitored 
with 
better than 5\% accuracy. The  momentum resolution of muons from the 
charged current quasi-elastic interactions~(CCQE) should be less than 10\%, and the 
threshold for 
detection of recoil protons  is required to be about 200~MeV/c. The 
$\nu_e$ fraction should be measured with an uncertainty of $\leq 10$\%.
A measurement of  the neutrino  beam direction, with a precision much better 
than 1 mrad, is required from the on-axis detector.  

\section{Near neutrino detectors}
\label{sec:nd280}
\subsection{On-axis neutrino monitor}
\begin{figure}[h]
\centering\includegraphics[width=13cm,angle=0]{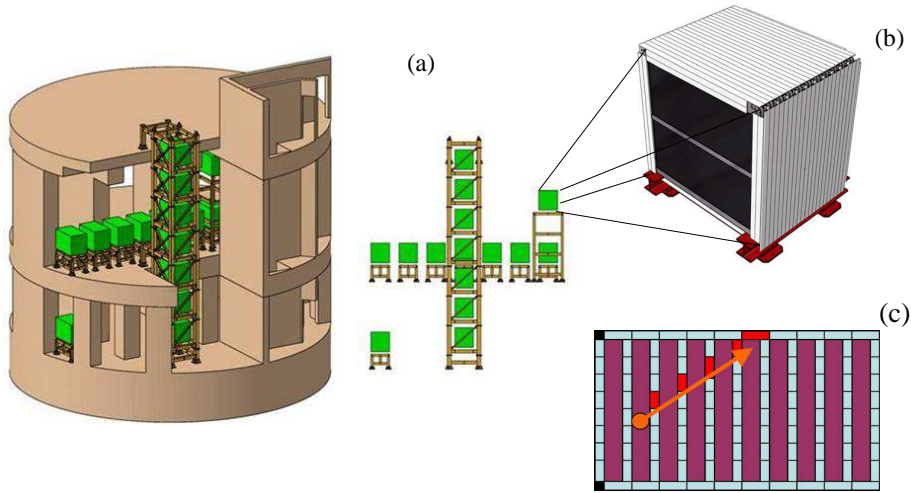}   
\caption{(a) schematic view of INGRID; (b) segmented iron-scintillator 
sandwich module; (c) a  charged current  neutrino interaction with muon 
track.}
\label{fig:ingrid}
\end{figure}

The role of the  on-axis neutrino detector~(INGRID) is to monitor the neutrino 
beam direction and profile on a day to day basis. It consists of $7 + 7$ 
identical modules, arranged to form a cross-configuration, and 2 diagonal modules, 
 as shown in   Fig.~\ref{fig:ingrid}.
INGRID samples the neutrino beam profile with an area of $9\times 9$ m$^2$.
Each  iron-scintillator sandwich module covers an area of $125 \times 125$ cm$^2$ 
and weighs 10 tons. The module consists of ten 6.5 cm-thick  
iron layers and 11
scintillator tracking  planes, and is surrounded by four veto counters. Each
tracking plane has one vertical and one horizontal scintillator layer 
composed  of $5\times 1\times
121$ cm$^3$  scintillator slabs.
Each scintillator has a central hole to insert a wavelength shifting~(WLS) 
fiber for light 
collection and routing to a photosensor. 
 The total mass of INGRID is $\sim 160$ tons.
A typical  event rate  detected by the center module  every spill is expected 
to  be  about 0.5 per ton, i.e. the whole INGRID will detect more  than $10^5$
neutrino events/day. In order to minimize the systematic from the uncertainty 
of the off-axis angle, the neutrino beam direction will be monitored by 
INGRID with a precision of $<< 1$ mrad each day at designed intensity.
 
\subsection{Off-axis near detector}
\begin{figure}[h]
\centering\includegraphics[width=10cm,angle=0]{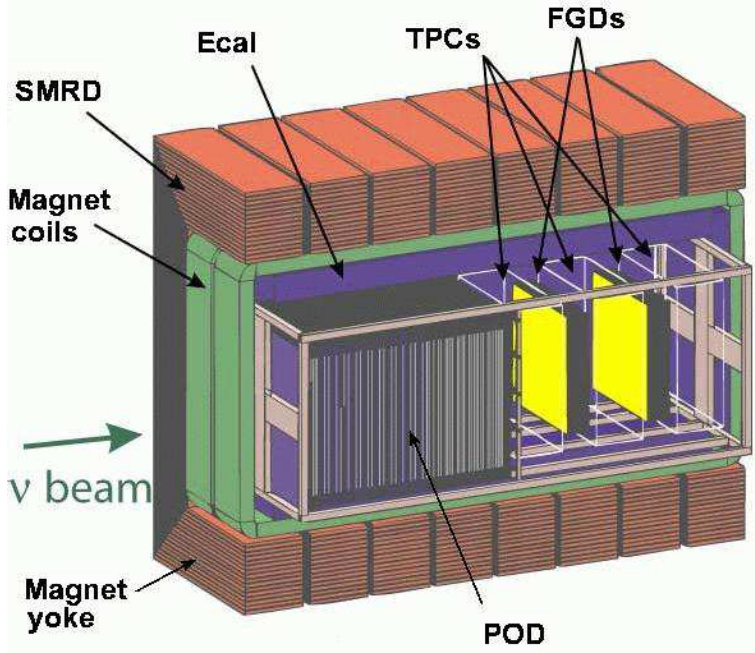}   
\caption{The cutaway view of the T2K  near detector.}
\label{fig:nd280}
\end{figure}

The off-axis detector (Fig.~\ref{fig:nd280})
 includes the UA1 
magnet operating with a magnetic field of 0.2 T, 
a Pi-Zero detector (POD), a tracking detector which includes  time projection 
chambers (TPC's) and fine grained scintillator detectors (FGD's), an 
electromagnetic calorimeter
(Ecal), and a side muon range detector~(SMRD).   

 \subsubsection{Photosensors}
 Wavelength 
shifting  fibers will be widely used for  readout of all 
scintillator detectors which are the main active element of  the ND280  
detector.  
 A magnetic field environment and  limited  space inside the 
UA1 magnet put serious constraints for the usage of standard photodetectors 
such as traditional multi-anode photomultipliers. Since the ND280 has about 
60k individual readout channels, the cost of 
photosensors is also very important. 
 After studying several candidates, 
a multi-pixel avalanche photo-diode operating in the 
limited Geiger multiplication mode was selected as the baseline detector.  These 
 novel devices  
are compact, well matched to spectral emission of WLS fibers, and insensitive 
to magnetic fields~\cite{gm1,gm2,andreev}. The required  parameters for these
photosensors from all ND280 subdetectors can be summarized as follows: 
an active area diameter of $\sim 1$ mm, photon detection efficiency for green 
light
$\geq 20$\% , pixels number $> 400$, and a dark rate at operating 
conditions $\leq 1$ 
MHz. The gain should be  $(0.5-1.0)\times 10^6$, 
the cross$-$talk $\sim10$\%, and 
pulse width  should be less than 100 ns to meet  the spill structure of the 
JPARC proton beam.  For 
calibration and control  purposes, it is very desirable  to obtain well separated  
single photoelectron peaks in amplitude spectra  at operating temperature.

\begin{figure}[h]
\begin{center}
\includegraphics[width=9cm,angle=0]{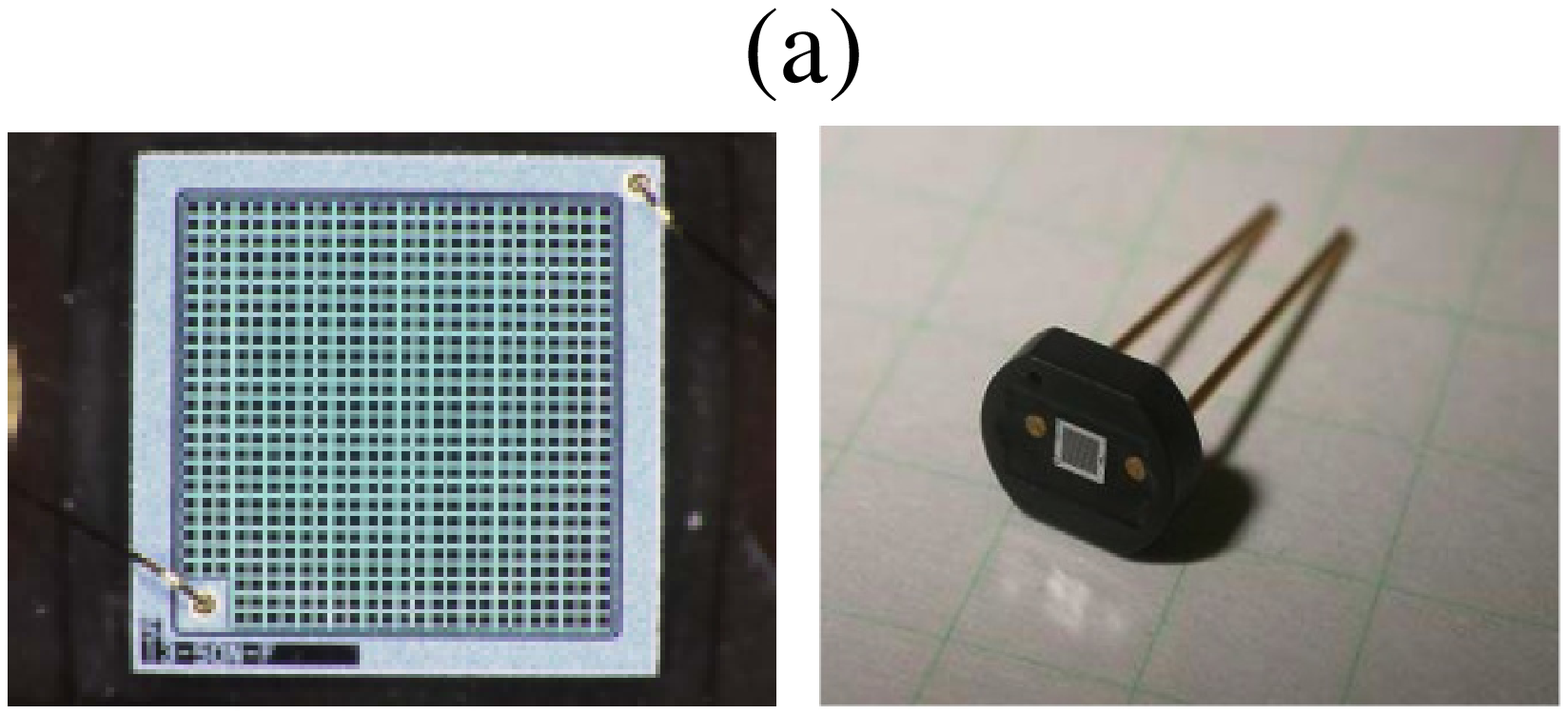}
\includegraphics[width=6cm,angle=0]{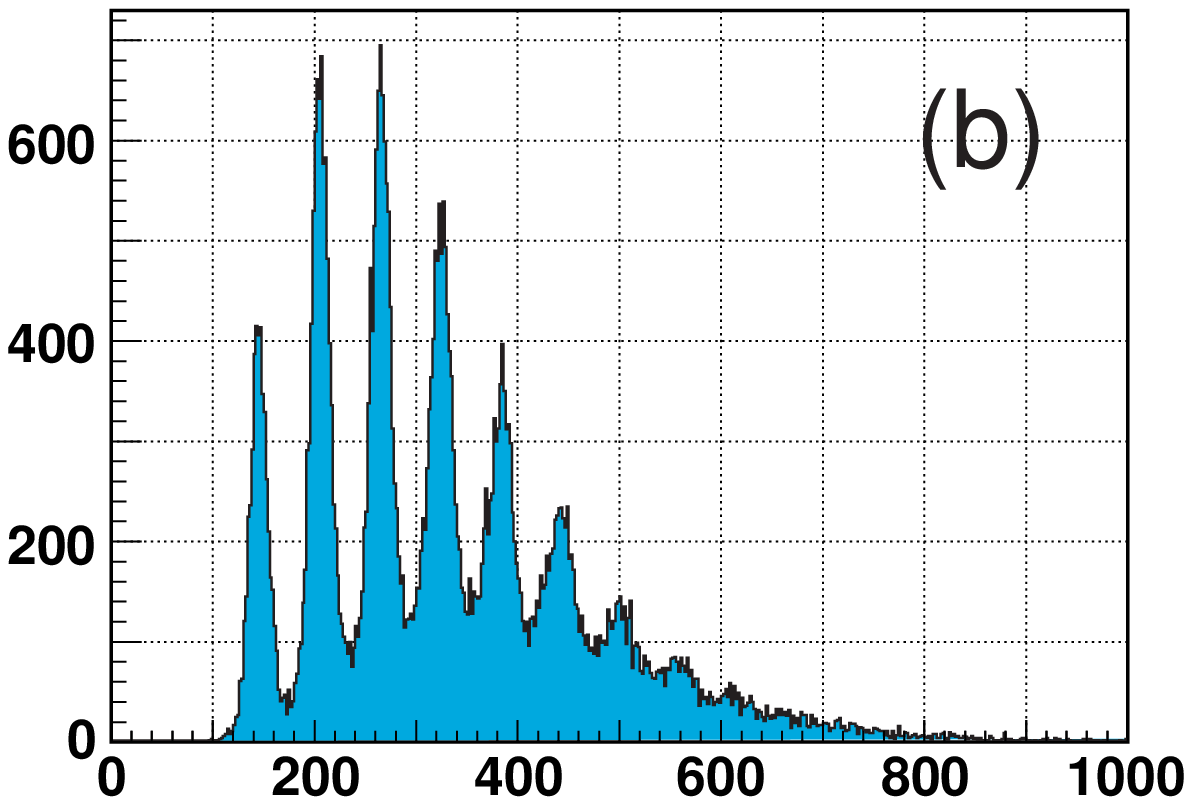}
\end{center}
\caption{(a) The photograph of a 667 pixel MPPC:  a magnified face view of 
an MPPC  with a sensitive area of $1.3\times 1.3$ mm$^2$ (left),
the package of this MPPC (right); (b) ADC spectrum from an LED 
signal. Clearly separated peaks at equal intervals correspond to 0, 1, 2, 
3... fired pixels.}
\label{fig:mppc}
\end{figure}

After a R\&D study of 3 years,  a Hamamatsu  MPPC was  chosen  as the 
photosensor for ND280. The description of this device
and its parameters can be found in Ref.~\cite{hamamatsu}. 
The final T2K version is a 667 pixel MPPC with a sensitive area 
of $1.3\times 1.3$ mm$^2$ (Fig.~\ref{fig:mppc}). 
 These devices
demonstrated  good performance at room temperature: a low cross-talk  value of about 10\%, 
a photon detection 
efficiency for green light of $\geq 25$\%,   a low dark rate of $\sim 0.3$ 
MHz at the operating voltage, a high gain of about 
 $0.7\times 10^6$, and a pulse width of less than 50 ns.  
 
\subsubsection{POD}
\begin{figure}[h]
\begin{center}
\includegraphics[width=6.2cm,angle=0]{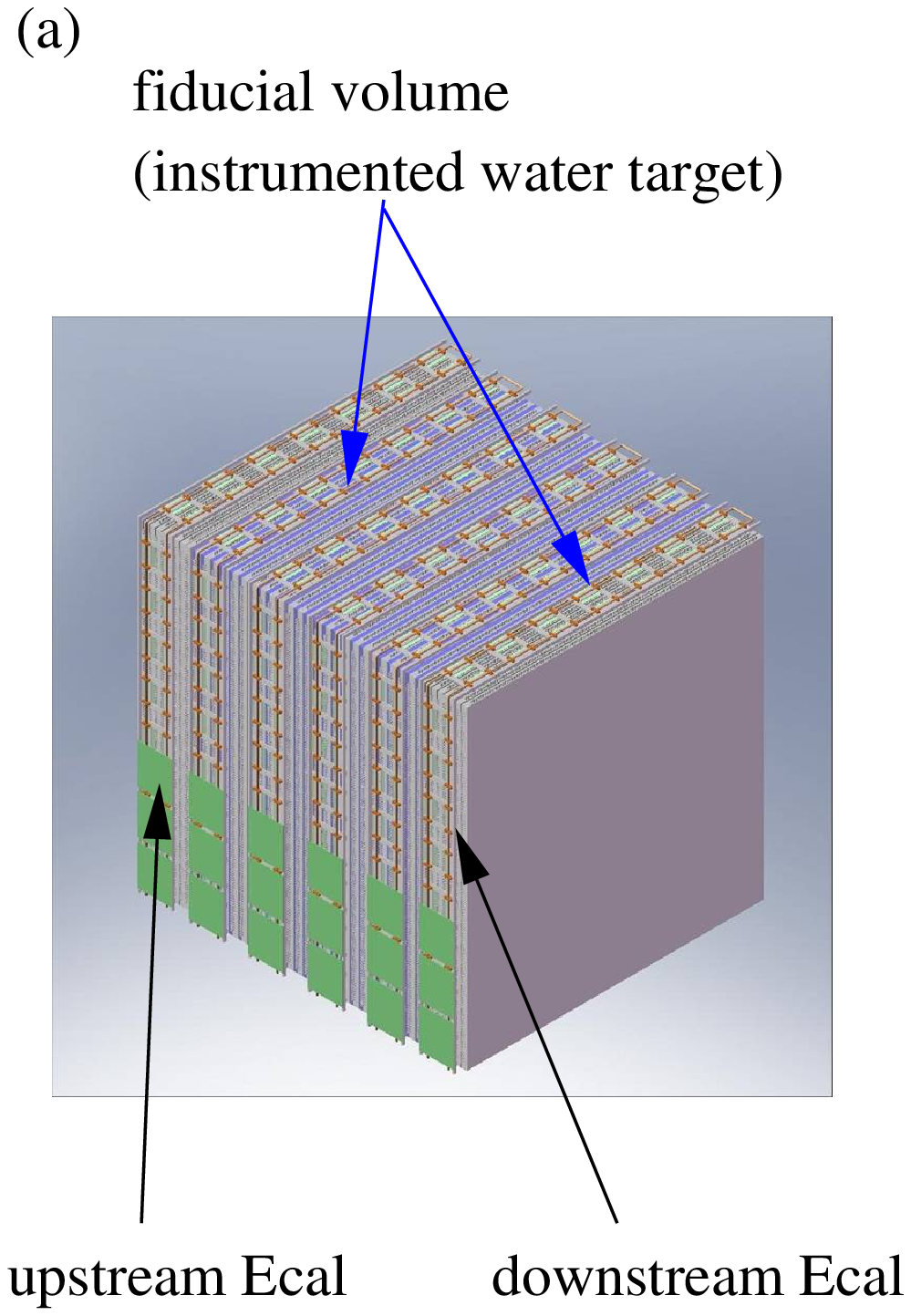}
\includegraphics[width=7.5cm,angle=0]{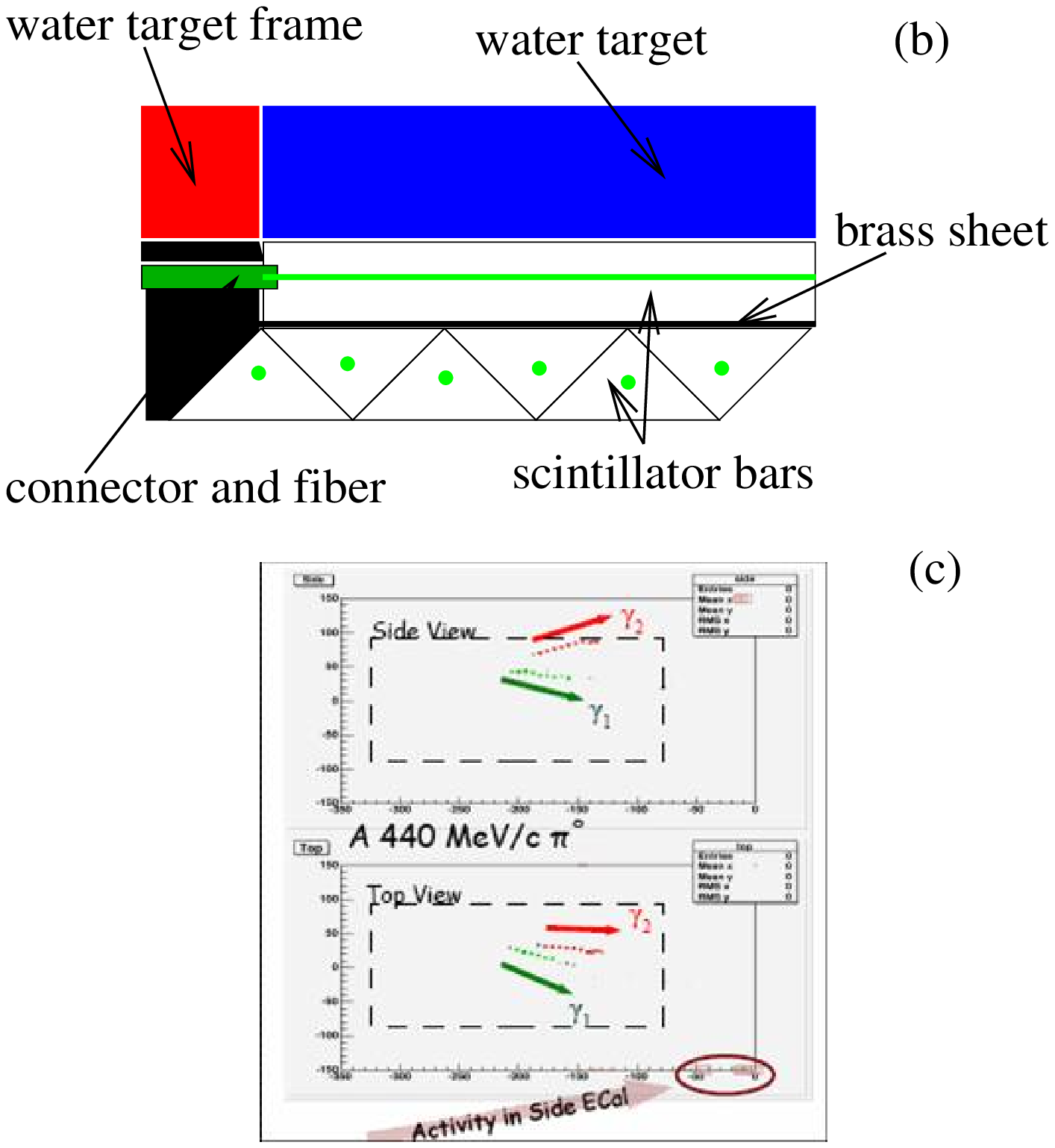}
\end{center}
\caption{(a) POD schematic view: the central region is constructed of
alternating water target and scintillator tracking layers; (b) one layer of 
the target region; (c) a  neutral current
$\pi^0$ event in POD. The horizontal and vertical axes are in centimeters. }
\label{fig:pod}
\end{figure}

The POD is optimized for measurement of the inclusive  $\pi^0$ production by  
$\nu_{\mu}$ interactions on oxygen and  will be installed in the upstream end of the 
magnet. The cross section measurements on an oxygen target will be achieved by 
using the following POD geometry: the upstream and downstream regions are
configured as electromagnetic calorimeters providing energy containment and
active veto, and the central region of the POD provides the fiducial mass for 
the $\pi^0$ measurements (Fig.~\ref{fig:pod}(a)).
 The schematic view of a plane of the POD target is shown in 
 Fig.~\ref{fig:pod}(b).
It consists of alternating water target layers of about 3 cm-thick  and 
tracking layers composed of X-Y extruded triangular shaped scintillator bars of 
17 mm in height and 32.5 mm at the base and a central hole for a WLS fiber.  
 A thin sheet of brass
($\sim 1.6$ mm-thick) is sandwiched in the 26 X-Y tracking layers of the target
region. The  upstream and downstream regions have 7 X-Y scintillator layers 
with 4 mm-thick lead radiators between them.
  The POD has a total mass 
of approximately 17 tons with a fiducial mass of about 3 tons of water and 
8 tons of other materials. 
 The tests of light yield of the POD
scintillators with MPPC's connected to one end of a Y11 WLS fiber 
 showed good results. The light yields for a minimum 
ionizing particles~(MIP) are 19.8 p.e./MeV and 8.7 p.e./MeV at 25 and 205 cm  
from the MPPC, respectively. With the mirrored  far end of the WLS fiber, 
the light yield 
of 15.7 p.e./MeV is much greater than the 5 p.e./MeV required for 
efficient reconstruction of electromagnetic showers.

Oxygen cross
section measurements will be made by comparing the interaction rate of events 
collected with water in the target region  versus similar running 
periods with water removed from the target region. A typical neutral current 
event with  a $\pi^0$ is shown in Fig.~\ref{fig:pod}(c) in which a neutral pion is
accompanied by a neutron.
The energy resolution for events fully contained in the active
target  is expected to be about 10\% + 3.5\%/$\sqrt{{\rm GeV}}$ and 
the reconstruction efficiency of a $\pi^0$ with a momentum $\geq 200$ MeV/c
is expected to be approximately 33\%.

\subsubsection{ND280 tracker}
The ND280 tracker consists of three TPC's and two FGD's, as shown in 
Fig.~~\ref{fig:nd280}. Its main function is 
to measure the  muon and electron  neutrino beam fluxes and  energy spectra, 
plus
various 
charged current cross sections.  The tracker is designed to  accurately 
measure CCQE events, the main 
process  at the T2K peak neutrino energy, 
\begin{equation}
\nu_{\mu} + n \to \mu^- + p.
\label{eq:ccqe}
\end{equation}
 In order to measure this process the reconstruction of both
 proton and muon  is useful. The proton will be primarily identified and measured 
 by the FGD while the muon will be measured by the TPC.  The initial neutrino energy 
 will be reconstructed from the muon momentum. The  measurements of CCQE events 
 will be used for flux normalization in the  oscillation analysis. \\
 
{\it FGD.} The ND280 will 
contain two FGD's, each 
with dimensions $1.84\times 1.84\times 0.3$ m$^3$ resulting in a  total target 
mass of about 2.0 tons.  
The first FGD will be an active scintillator detector, similar to the 
SciBar detector~\cite{scibar} of the K2K experiment. 
It consists of thirty scintillator layers of 192  
$0.96\times 0.96 \times 184$ cm$^3$ extruded  
scintillator bars which are   arranged 
in alternating vertical and horizontal layers perpendicular to the beam
direction.  The second FGD is composed of seven X-Y 
sandwiches  of scintillator 
layers alternating with six 3-cm thick layers of water. The weight of the  
scintillator is 0.56 ton and  of water, 0.44 ton. The readout of each
scintillator bar  is provided by an MPPC connected to one end of a WLS fiber 
inserted in 
a central hole. 
A beam test of scintillator bars performed at TRIUMF showed that the light yield
for 120 MeV/c pions, muons, electrons will be more than 10 p.e. at the far end 
from a photosensor without mirroring the far end of the fiber. The  
mirroring  increases the light yield by $\geq 80$\%,
guaranteeing a detection efficiency of more 
than 99\% for  minimum ionizing particles. The FGD  allows a clear
separation between protons and pions using dE/dx information and tagging 
Michel electrons from the decay of short-ranged pions.
Comparing the interaction rates in both FGD's permits separate measurement 
of neutrino cross sections on carbon and on water.
About $4\times 10^5$ 
neutrino interactions are expected in both FGD modules for a one year exposure 
with $10^{21}$ protons on target.

{\it TPC.} The primary purpose  of the TPC   is 
to measure the 3-momenta of muons produced in  CCQE interactions in the FGD. 
 The TPC will use a 
low diffusion gas  to obtain the 
momentum resolution of $\leq 10$\% for particles below 1 GeV/c. A 700 $\mu$m 
space point resolution per ``row'' of pads  is required to achieve this momentum resolution. 
The absolute energy scale will be checked at the 2\% level using the invariant 
mass of neutral kaons produced in DIS neutrino interactions and decaying in the
TPC volume. A good dE/dx resolution of $< 10$\% 
is expected for 72 cm long tracks which will provide better than 5$\sigma$ 
separation between muon and electron tracks 
in the momenta range 0.3-1.0 GeV/c. 

\begin{figure}[h]
\centering\includegraphics[width=8cm,angle=0]{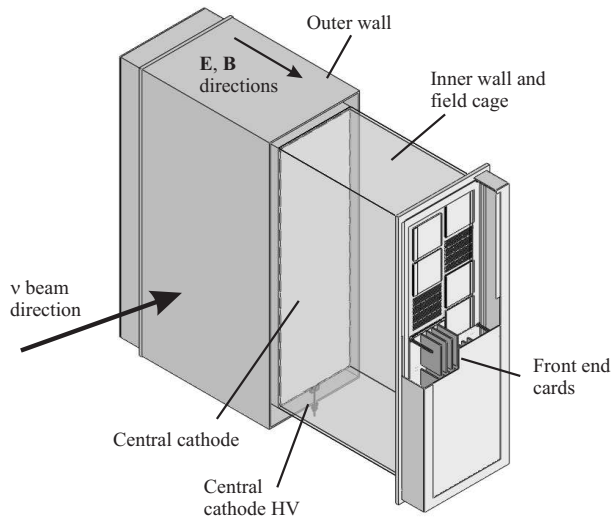}   
\caption{The layout of TPC showing the inner and outer boxes and the 
central cathode.}
\label{fig:tpc}
\end{figure}

The  three TPC modules  are 
rectangular boxes with outer 
dimensions of approximately $2.5\times 2.5$ m$^2$ in the plane perpendicular 
to the neutrino beam direction, and 0.9 m along the beam direction. A simplified
drawing of the TPC is shown in Fig.~\ref{fig:tpc}.
The TPC modules are operated at an electric field of 200 V/cm. 
The central cathode, which divides the drift space into two halves to limit the 
maximum drift distance to $\sim 1$ m, will be  at a potential of -25 kV. The
baseline gas choice is Ar(95\%)--CF$_4$(3\%)--iC$_4$H$_{10}$(2\%).

The `bulk' Micromegas detectors will be used to instrument the TPC readout
plane. The active surface area  of the  
Micromegas 
module is  $359\times 342$ mm$^2$ with 1726 active pads of $9.7\times 6.9$ 
mm$^2$.  12  Micromegas modules   will be used for each
readout plane of the TPC.
 In total, the 3 TPC's will consist of 72 modules with  $\sim 124000$ readout 
channels. 

The first prototypes of Micromegas detectors have been tested with cosmic 
muons in the former HARP 
field cage setup with a magnetic field~\cite{micromegas}  and demonstrated good
momentum resolution of 8.3\% at 1 GeV/c. A dE/dx resolution of 
about 12\%  for track lengthes of about 40 cm and a good 
uniformity of $\sigma = 3.4$\% for  the gain $\sim 1000$  have been obtained. 

\begin{figure}[h]
\centering\includegraphics[width=9cm,angle=0]{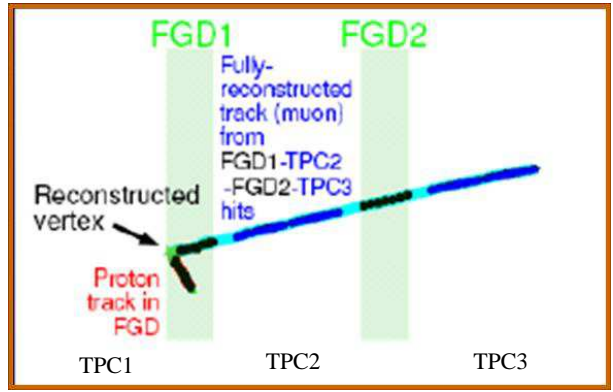}   
\caption{Typical CCQE event in the tracker.}
\label{fig:ccqe_track}
\end{figure}

A typical CCQE event for neutrino interaction in FGD1 is shown in 
Fig.~\ref{fig:ccqe_track}.
The   reconstruction efficiency of CCQE 
events produced in the FGD with a track in the TPC is estimated to be about 
50\% at a $E_{\nu} \sim 0.7$ GeV.

\subsubsection{Electromagnetic calorimeter}
\begin{figure}[h]
\centering\includegraphics[width=7cm,angle=0]{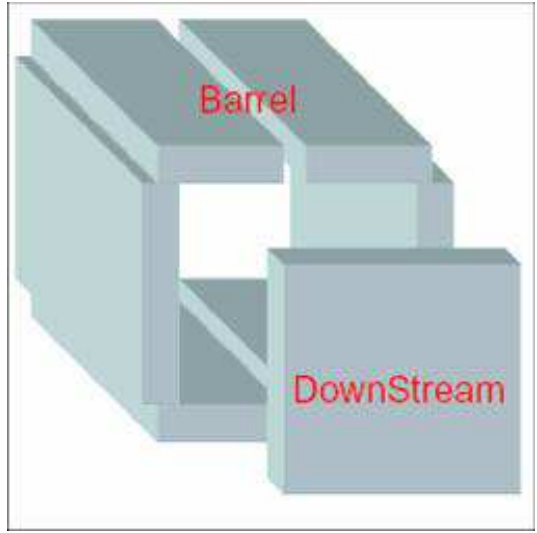}   
\caption{Basic structure of the electromagnetic calorimeter.}
\label{fig:ecal}
\end{figure}

The Electromagnetic calorimeter (Ecal) shown in Fig.~\ref{fig:ecal} 
consists of two sections. 
One surrounds the POD (POD Ecal) for detectioning photons and muons escaping the 
POD, and the second section, surrounding the FGD's and TPC's  (TEcal), detects 
particles leaving the tracking volume. 

TEcal modules are made of 4 cm-wide, 1 cm-thick plastic
scintillator bars arranged in 32 layers and separated by 31 layers of 1.75
mm-thick lead sheets.  The orientation of the bars alternates between layers 
so that the bars in any layer are perpendicular to the bars in the two adjacent 
ones. This bar width allows good $\pi^0$ reconstruction efficiency and provides 
the spatial resolution required for reconstruction  of the direction  of detected
photons. The active length of the TEcal along the neutrino beam is 384 cm and  
the total depth   is 50 cm corresponding to 10.5$X_0$. TEcal has two side
modules, one on each side of the UA1 iron yokes, one top and one 
bottom modules, each is split into two (left and right) so that they can move 
with the magnet yoke when the magnet opens.
All scintillator 
bars have a hole in the center with a 1 mm WLS fiber inserted in it. All 
long bars running along the neutrino beam are readout by an MPPC at each end
(double-end readout), while all shorter bars (perpendicular to the neutrino
beam) are mirrored at one end and readout by an MPPC at the other (single-end
readout). The downstream Ecal is a single module with the same granularity 
as TEcal modules with an effective depth of 11$X_0$. It is located at the 
downstream end of the magnet and covers 
an active surface area of $2\times 2$ m$^2$. All bars have double-ended readout. 
The total weight of the TEcal and downstream Ecal is 28.3 tons. 

 The POD Ecal has modules with coarser segmentation and less total $X_0$ and 
 does not provide good energy and spatial resolution required for $\pi^0$ 
 reconstruction. These modules consist of 6 scintillator layers separated by 
 5 layers of 5 mm-thick lead converters resulting in an effective depth of 
 $4.5X_0$. 
 
 The energy resolution of TEcal, dominated by sampling fluctuations, is
 estimated to be about 7.5\%/$\sqrt{E(\rm GeV)}$ for energies up to 
 5 GeV.  TEcal is expected to provide good electron/pion separation. An
 efficiency of 90\% for electrons is expected with 95\% pion rejection. 
 
\subsubsection{Side muon range detector}
Muons which escape at large angles with respect to the neutrino 
beam can not be measured by the TPC's. However, they will intersect in the iron yoke 
and therefore a muon's momentum can be obtained from its range by 
instrumenting the iron at various depths.  About 40\% of muons from 
CCQE reactions and about 15\% of muons from  
charge current 
non-quasi-elastic reactions  are expected to enter the SMRD.  In addition, 
the SMRD will be used to veto events from 
neutrino interactions in the magnet and  in walls of the ND280 pit and will 
provide a cosmic trigger 
for calibration of inner detectors. 

\begin{figure}[h]
\centering\includegraphics[width=12cm,angle=0]{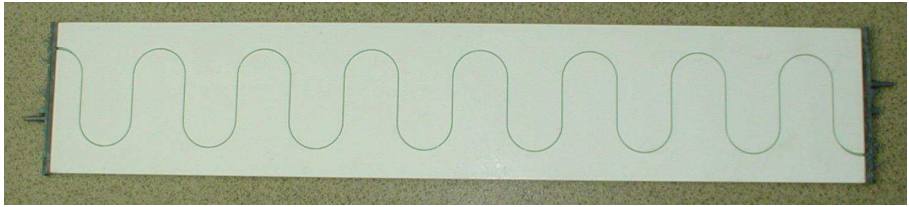}   
\caption{The  SMRD counter with embedded Kuraray Y11 WLS fiber.}
\label{fig:smrd_counter}
\end{figure}

\begin{figure}[h]
\centering\includegraphics[width=11cm,angle=0]{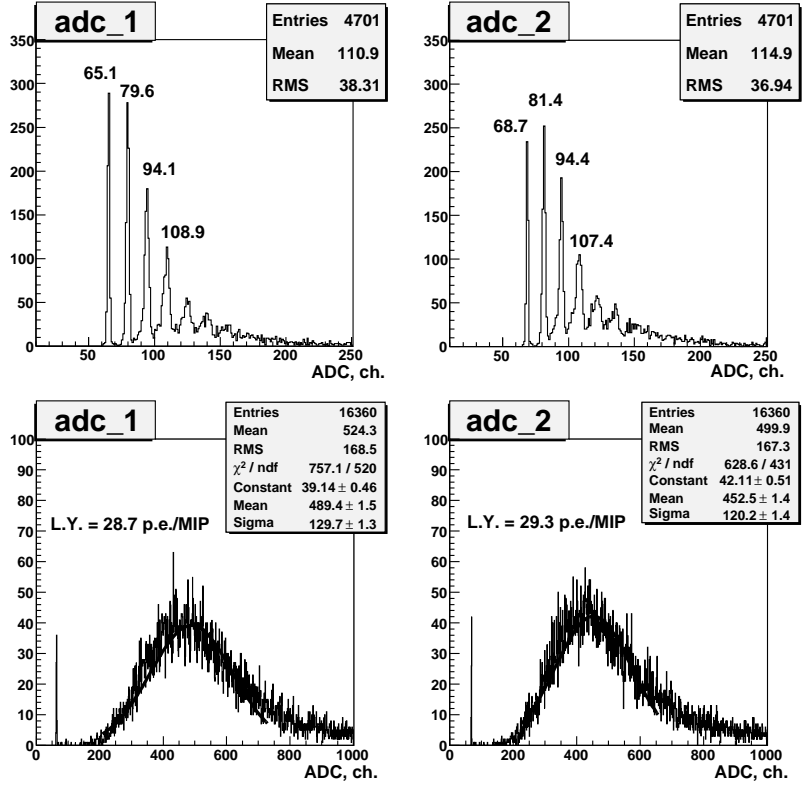}   
\caption{The ADC spectra of the SMRD counter  for  minimum ionizing 
particles measured at 22$^{\circ}$C. The light yield (sum of both ends) is 
equal to 58 p.e.}
\label{fig:adc_smrd_spectra}
\end{figure}

The UA1 iron yoke  
consists of 16 C-shaped elements made of sixteen 5 cm thick iron plates, with 
1.7 cm air gap between the plates and is segmented in 12 azimuthal sections. 
The active component of the SMRD will consist of 0.7 cm thick scintillator slabs
sandwiched between the iron plates of the magnet yokes. 
Details of the extrusion of the scintillator slabs  and the method  of 
etching the plastic surface by a chemical agent   can be found 
in Ref.~\cite{extrusion}. For the readout, we  employ a 
single WLS fiber embedded in a serpentine shaped (S--shape)
groove, as shown in Fig.~\ref{fig:smrd_counter}.
  Such a 
shape allows the fiber to collect the scintillation light over the whole 
surface of a scintillator slab~\cite{smrd_nim}. Two MPPC's are  coupled to 
both ends of a  
WLS fiber glued into the S--shape 
groove. The detector performance has been tested using cosmic muons. Typical 
ADC spectra for  
MIP's obtained with $1.0\times 1.0$ mm$^2$ MPPC's are  
shown in Fig.~\ref{fig:adc_smrd_spectra}. 
The SMRD counters tests  resulted in a high detection efficiency measurement 
of 
greater than 99\%, a time resolution of about 1 ns, and a spatial resolution 
along the slab of about 8 cm   for minimum ionizing particles.
\section{Conclusion}
The T2K experiment has  a rich physics potential  and provides an excellent  
opportunity to greatly extend our  understanding of neutrino properties.
To achieve the physics goals of T2K, the complex of near neutrino detectors  
needed for measurement of the unoscillated neutrino beam properties 
is under construction. The on-axis detector will be ready  to accept the first 
neutrino beam in April 2009, the installation and commissioning  of the whole 
off-axis detector  will be finished during 2009. The T2K experiment is 
expected to start data taking in 2009.  
 
{\bf Acknowledgments.} I thank M.~Gonin, A.~Grant,  D.~Karlen, T.~Nakaya, 
V.~Paolone and  M.~Yokoyama for providing material for the talk and useful 
comments on the manuscript.  This work was supported in part by  
the ``Neutrino Physics'' Program 
 of the Russian Academy of  Sciences and by the RFBR (Russia)/JSPS (Japan) 
 grant \#08-02-91206.



\end{document}